\documentclass[12pt,preprint]{aastex}

\shorttitle{Shocks and PDRs in Galactic center}
\shortauthors{Mart\'{i}n et al.}

\begin{document}

\title{Tracing shocks and photodissociation in the Galactic center region}

\author{Sergio Mart\'{i}n}
\affil{Harvard-Smithsonian Center for Astrophysics, 60 Garden St.,  02138, Cambridge, MA, USA}
\email{smartin@cfa.harvard.edu}

\author{M.A. Requena-Torres}
\affil{Max-Plack-Institut f\"ur Radioastronomie, Auf dem H\"ugel 69, D-53121 Bonn, Germany}
\author{J. Mart\'{i}n-Pintado}
\affil{Departamento de Astrofis\'{i}ca Molecular e Infrarroja, Instituto de Estructura de la Materia, CSIC, Serrano 121, E-28006 Madrid, Spain}
\author{R. Mauersberger}
\affil{Instituto de Radioastronom\'{i}a Milim\'etrica, Avenida Divina Pastora 7, Local 20, E-18012 Granada, Spain}

\begin{abstract}
We present a systematic study of the HNCO, C$^{18}$O, $^{13}$CS, and C$^{34}$S emission towards 13 selected molecular clouds in the Galactic center region
\footnote{Based on observations carried out with the IRAM 30-meter telescope. IRAM is supported by INSU/CNRS (France), MPG (Germany) and IGN (Spain).}.
The molecular emission in these positions are used as templates of the different physical and chemical processes claimed to be dominant in the 
circumnuclear molecular gas of galaxies.
The relative abundance of HNCO shows a variation of more than a factor of 20 among the observed sources.
The HNCO/$^{13}$CS abundance ratio is highly contrasted (up to a factor of 30) between the shielded molecular clouds mostly affected by shocks,
where HNCO is released to gas-phase from grain mantles, and those pervaded by an intense UV radiation field, where HNCO is photo-dissociated and CS production
favored via ion reactions.
We propose the relative HNCO to CS abundance ratio as a highly contrasted diagnostic tool to distinguish between the influence of shocks and/or
the radiation field in the nuclear regions of galaxies and their relation
to the evolutionary state of their nuclear star formation bursts.


\end{abstract}


\section{Introduction}

The central few hundred parsecs of many galaxies, including our own, harbor large amounts of molecular gas and are the sites of past, present or future
star formation.
The interaction of the gas with the gravitational potential in such areas, with the radiation and outflows from recently formed and evolved stars,
and with the central objects makes that the physical processes influencing this molecular gas are much more complex than in disk clouds.
These physical processes not only dominate the energy balance in the circumnuclear environment
of galaxies, but they also determine if stars could be formed within such clouds.
In the central 300 pc of our own galaxy we find giant molecular clouds (GMC) harboring
regions of massive star formation and heavily affected by shocks \citep{Pintado01}; 
large regions where molecular gas is pervaded by intense UV photodissociating radiation (PDRs) stemming from massive star clusters
\citep{Krabbe95,Figer99,Rodriguez01};
strong X-ray emission \citep[][and references therein]{Sidoli01} as well as the ultimate massive black hole candidate \citep{Eisen05,Ghez05}.
This variety of energetic processes not only modifies the distribution, kinematics, density and temperature of the gas but
also drives different types of chemistry in the molecular gas in the Central Molecular Zone \citep[CMZ, ][]{Morris96}.

Comparing the chemical composition between Galactic regions with known dominant heating mechanisms
can be used to trace the heating mechanisms in the central region of galaxies.
Unbiased spectral line surveys of selected regions within our Galaxy allow to identify the different chemical signatures which trace the 
different dominant heating mechanisms affecting the nuclear interstellar medium (ISM) in galaxies with different type of activity and/or in different
evolutionary states \citep{Martin06b}.

One of the molecules whose abundance appears to be a particularly unambiguous tracer of shock heating is HNCO.
This asymmetric rotor has been observed in high mass molecular hot cores with fractional abundances relative to H$_2$ of $\sim10^{-9}$ \citep{Zinchen00} 
as well as in translucent and dark clouds with similar abundances \citep{Turner99}.
The observations of HNCO toward the photodissociation regions in the Orion bar shows
abundances of $\lesssim$1.6$\times$10$^{-11}$ \citep{Jansen95}, indicating  this molecule is easily destroyed by the UV radiation.
Within the Galactic center region, the low resolution C$^{18}$O (which is thought to be a tracer of the total H$_2$ column density)
and HNCO maps by \citet{Dahmen97} as well as the close up view towards the Sgr\,A region 
by \citet{Lindqvist95} clearly show differences in their distribution which can not be ascribed to only differences in excitation but to real 
differentiation in the chemical properties of the different molecular 
complexes in the Milky Way.
\cite{Minh06} have mapped the Sgr\,B2 region and they identified an expanding ring-like structure in the HNCO emission, clearly different from that of
$^{13}$CO and HCO$^+$.
This suggests that HNCO is likely enhanced by the shock interaction between the expanding ring and the main cloud.
Similar chemical differences in the distribution of the HNCO emission have also been observed at larger scales towards external galaxies \citep{Meier05}.
Therefore HNCO seems to be a potential good tracer of different heating mechanisms in the nuclei of galaxies. However, there is not a systematic 
study of the physical conditions and the relative abundances of HNCO in a sample of molecular clouds in the CMZ affected
by different type of heating mechanisms.

In this paper we present systematic multitransition observations of HNCO toward 13 molecular clouds in the Galactic center which
are thought to be affected by different processes such as shocks, UV radiation, and X-rays.
We explore the potential of HNCO to be used as a chemical discriminator of the different physical processes affecting the chemistry 
of the ISM.
We show that the HNCO abundance ratio with respect to C$^{18}$O, $^{13}$CS, and C$^{34}$S is an excellent discriminator between the molecular clouds
affected by the UV radiation and moderate shocks.
The proposed CS vs HNCO diagnostic diagram can be used to trace these type of activities in the extragalactic ISM.

\section{Observations and Results}

The observations were carried out with the IRAM 30\,m telescope on Pico Veleta, Spain.
The thirteen sources observed throughout the Galactic center region are shown in 
Table~\ref{tab:sources}, where both the coordinates and the Galactic center complex where each source is located are given.
During the observation we switched the position of the telescope between the source and a reference every 60 seconds. As reference positions
we used $l=0.65,\,b=0.2$ ($\alpha_{\rm J2000}=17^{\rm h}46^{\rm m}23\fs0, \delta_{\rm J2000}=-28\arcdeg16\arcmin37\arcsec$) for the sources in the Sgr\,B 
complex and $l=-0.25,\,b=-0.25$ ($\alpha_{\rm J2000}=17^{\rm h}46^{\rm m}00\fs1, \delta_{\rm J2000}=-29\arcdeg16\arcmin47\arcsec$) for those around Sgr\,A 
in order to minimize the distances between source and reference for the sake of baseline stability.
No significant emission was found in these reference positions when previously observed in frequency switched mode tuned to the 
frequencies of the CS $J=3-2$ and $J=5-4$ transitions at 146.9 and 244.9\,GHz, respectively.

We observed the transitions of C$^{18}$O, C$^{34}$S, $^{13}$CS and HNCO indicated in Tables~\ref{tab:lines} and ~\ref{tab:lines2}.
We have also observed the HNC$^{18}$O J=7$_{0,7}$--6$_{0,6}$ line at 145.5\,GHz in two positions with large HNCO column densities to estimate
the opacity of the HNCO line.
The SIS receivers were tuned to SSB with image band rejections typically larger than 10\,dB.
At the observed frequencies the beamwidths of the telescope were $22''$ (3\,mm), $16''$ (2\,mm) and $10''$ (1.3\,mm).
As spectrometers we used the $512\times1$\,MHz filter bank for the 3\,mm transition and the $256\times4$\,MHz filter banks
for the 2 and 1.3\,mm lines.

\begin{figure}
\epsscale{1.3}
\includegraphics[angle=0,scale=0.75]{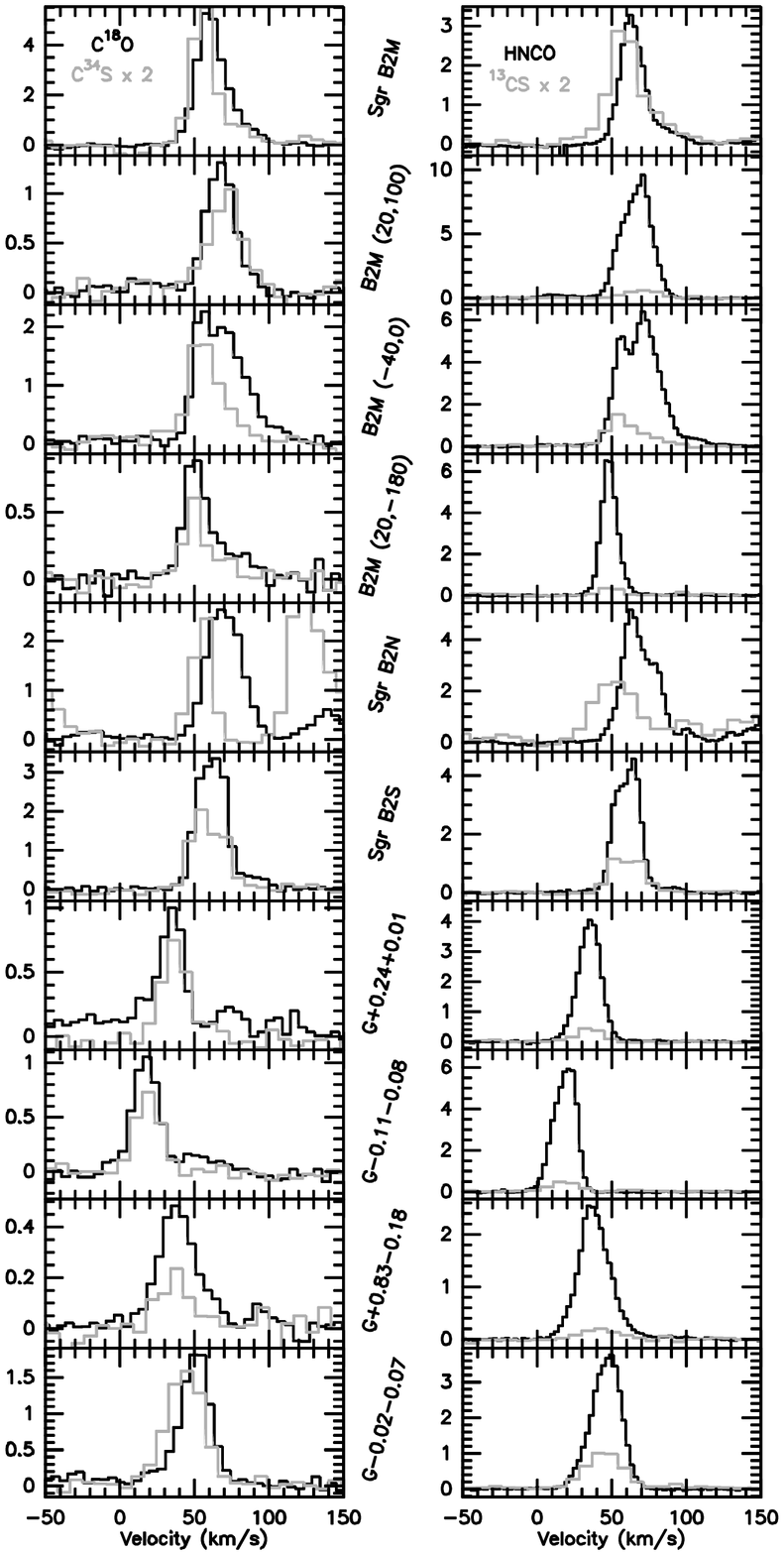}
\includegraphics[angle=0,scale=0.75]{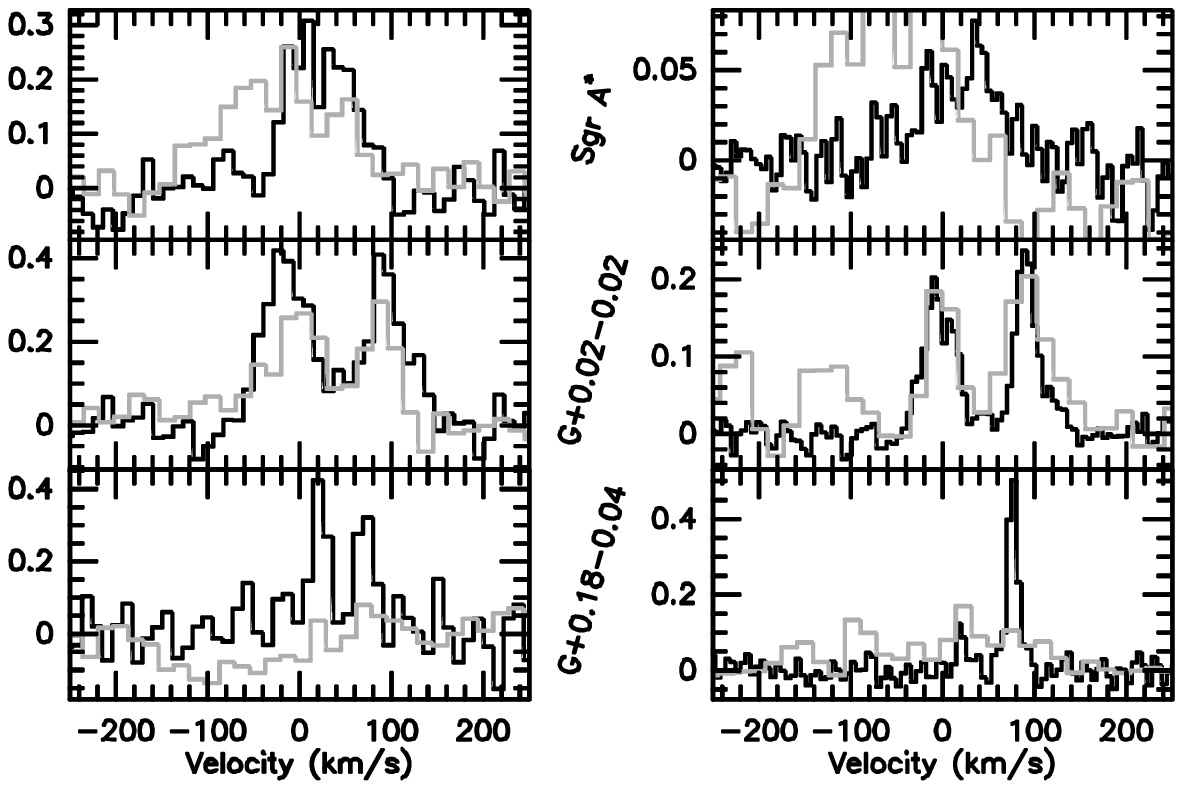}
\caption{Sample spectra of the brightest transitions of each observed molecule, namely $\rm C^{18}O(2-1)$, $\rm C^{34}S(3-2)$,
$\rm HNCO(5_{0,5}-4_{0,4})$ and $\rm ^{13}CS(3-2)$. The source identification is shown between the two spectra. 
The intensity of $\rm C^{34}S(3-2)$ and $\rm ^{13}CS(3-2)$ has been multiplied by a factor of two for the sake of clarity.
Temperature scale is $T_{\rm A}^*\,(\rm K)$.}
\label{fig:spectra}
\end{figure}

Fig~\ref{fig:spectra}
show a sample of spectra where the brightest among the observed transitions of each species have been selected for each source.
While single Gaussian profiles were fitted to most of the observed transitions, two Gaussians have been fitted to those sources showing asymmetric line profiles. 
The resulting integrated intensities and the radial velocities derived from the Gaussian fits are shown in 
Table~\ref{tab:lines} for the C$^{18}$O, C$^{34}$S and $^{13}$CS transitions and in Table~\ref{tab:lines2} for those of HNCO.

\begin{figure}
\includegraphics[angle=0, width=0.9\textwidth]{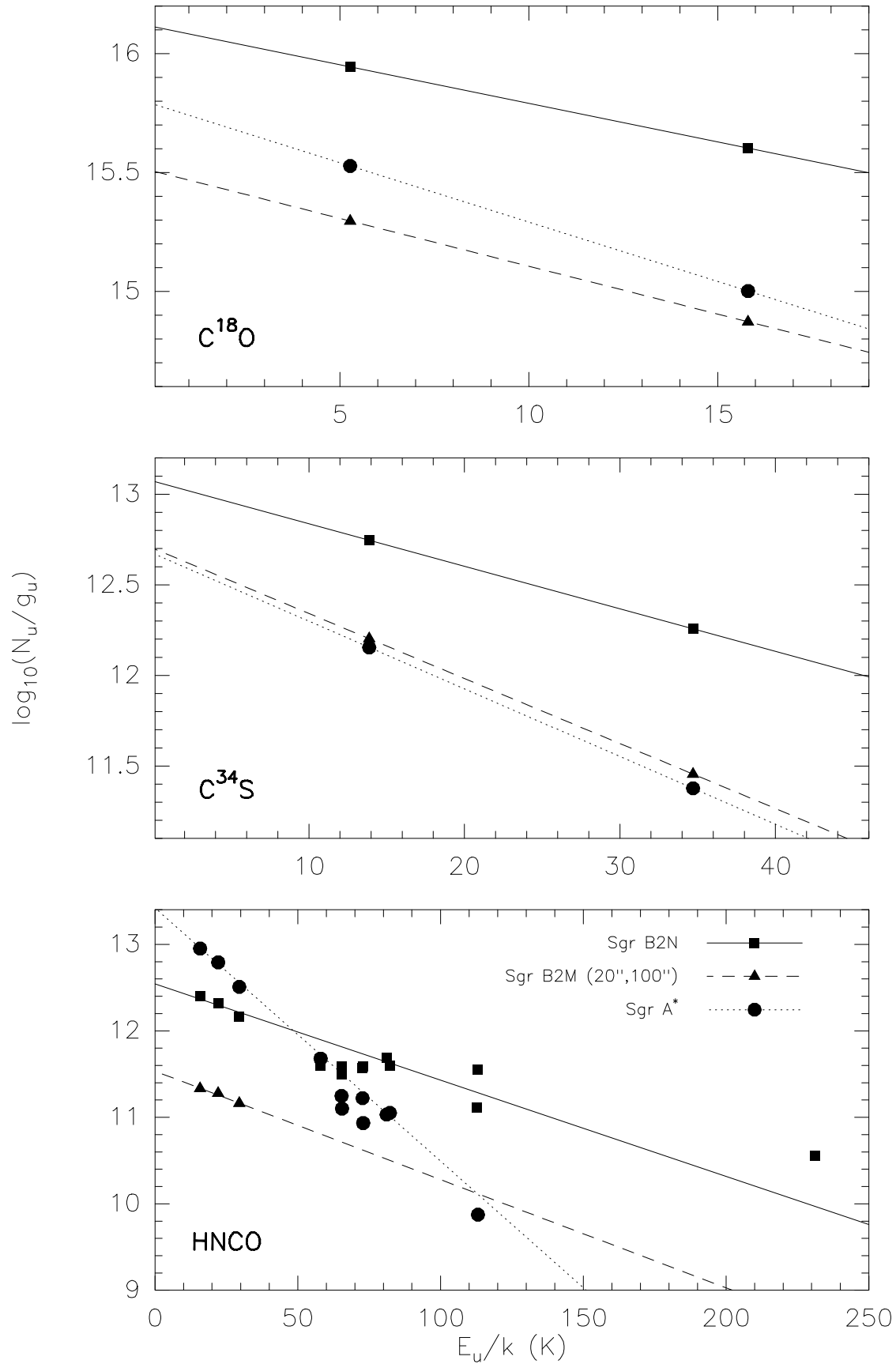}
\caption{Sample of rotational diagrams derived from C$^{18}$O, C$^{34}$S and HNCO in three selected representative sources.}
\label{fig:rotational}
\end{figure}

We have derived the column density and excitation temperature by assuming that the excitation of all molecules is in local thermodynamic equilibrium (LTE),
the lines are optically thin and the source is extended as compared to the telescope beam
\citep[see][for a detailed discussion on rotational diagrams]{Goldsmith99,Martin06b}.
We do not detect emission for the HNC$^{18}$O J=7$_{0,7}$--6$_{0,6}$ line to a limit of $<30$\,mK. 
Thus, for the typical $^{16}$O/$^{18}$O ratio of 250 \citep{Wilson94}, the observed HNCO lines appear to be optically thin.
In fact the $^{16}$O/$^{18}$O ratio we derived for the Sgr~B2
($-40, 0$) position is $>300$, larger than the canonical value of 250.
Fig~\ref{fig:rotational} shows the rotational diagrams derived for three typical sources,
namely a hot core, Sgr\,B2N, a GC molecular cloud, Sgr\,B2M($20'',100''$), and a
PDR in the circumnuclear disk (CND), Sgr\,A$^*$.
Rotational temperatures ($T_{\rm rot}$) derived from C$^{18}$O and C$^{34}$S are similar for all the sources, with average values
of $10.3\pm1.9$\,K and $12\pm3$\,K, respectively.
Thus, we assumed a $T_{\rm rot}=10$\,K to derive the $^{13}$CS column densities.
On the other hand, HNCO shows a wider range of rotational temperatures from 10\,K up to 56\,K. 
However, for most of the sources in the GC
we derive values of $T_{\rm rot}\sim$10\,K. 
Only the hot cores like Sgr\,B2\,M, N, and S and the gas associated with photo dissociation regions (see
Sec.~\ref{sect.PDRs}) show larger $T_{\rm rot}$. In fact, the HNCO diagrams of Sgr\,B2 N and M are fitted by a two gas 
component with temperatures of $\sim20$\,K, similar to the average found in the other sources, and $\sim100$\,K, typical of the hot cores.
From the obtained rotational temperatures we have derived column densities shown in Table~\ref{tab:coldens}.

\begin{figure}
\includegraphics[angle=-90, width=0.9\textwidth]{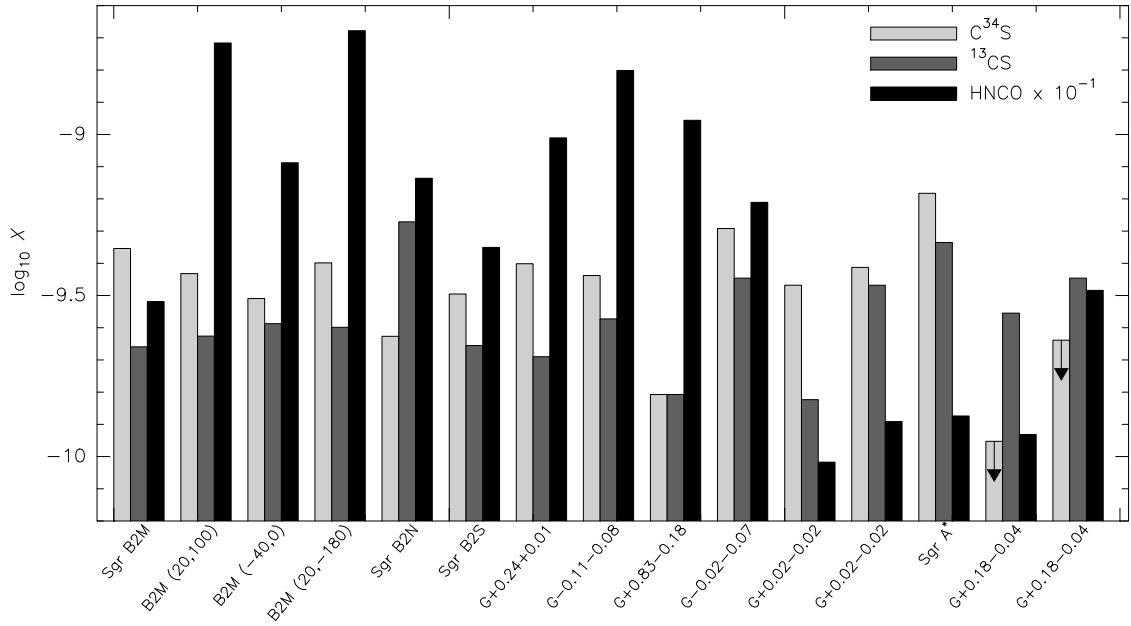}
\caption{Fractional abundances relative to H$_2$ assuming a conversion factor of $\rm C^{18}O/H_2=1.7\times10^{-7}$ \citep{Frerking82}.
Arrows represent upper limits to the observed abundances.}
\label{fig:abundances}
\end{figure}

Fig.~\ref{fig:abundances} shows the abundances of C$^{34}$S, $^{13}$CS and HNCO relative to H$_2$, where a C$^{18}$O/H$_2$ conversion factor of
$1.7\times10^{-7}$ has been assumed \citep{Frerking82}.
We note that the CS isotopologues show similar abundances for all the sources with changes smaller than a factor of 4 among all the observed sources. The C$^{34}$S/$^{13}$CS
ratio is also constant within a factor of 2.4 for all the sources except for the hot core Sgr\,B2N and G$+0.18-0.04$.
Nevertheless, the non-detection of C$^{34}$S in G$+0.18-0.04$ implies a surprisingly low abundance of this isotopical substitution in this source.
The upper limit to the C$^{34}$S/$^{13}$CS ratio of $\sim0.5$ is a factor of 3 below the average ratio derived for the other positions,
suggesting a significant underabundance of $^{34}$S in G$+0.18-0.04$  with respect to $^{13}$CS.

\section{Relative HNCO and $^{13}$CS abundances}


CS has been commonly used as a tracer of gas typically 100 times denser than that traced by CO
\citep[$n>10^4\,\rm cm^3$,][]{Mauers89a,Mauers89b,McQuinn02}.
The main precursor for the formation of CS in gas phase is S$^+$ \citep[through chain reactions forming CS$^+$ and HCS$^+$,][]{Drdla89}.
Therefore, even if photodissociation is the main mechanism of CS destruction, this molecule is observed to survive in regions with 
high UV radiation field, likely due to an enhancement of S$^+$ \citep[with predicted abundances up to $\sim 10^{-5}\,\rm cm^{-3}$,][]{Stern95}.
In fact, CS has been used to estimate the abundance of the observationally elusive S$^+$ in PDRs \citep{Goico06}.

HNCO may trace material even denser than CS \citep[$n_{\rm H_2}\gtrsim 10^6\rm cm^{-3}$;][]{Jackson84}.
The presence of HNCO in the icy grain mantles, though not directly observed, is suggested by the detection of OCN$^-$,
which is thought to be produced by the reaction on grains between HNCO and NH$_3$ \citep[][and references therein]{Novaz01}.
Though HNCO can be formed via gas phase reactions \citep{Iglesias77,Turner99}, its formation in solid phase seems to be more efficient \citep{Hasegawa93}.
The observed correlation of the HNCO emission with that of SiO in high mass star forming molecular cores suggests that HNCO abundance in 
gas phase is enhanced by grain erosion and/or disruption by shocks \citep{Zinchen00}. This correlation is also suggested for the Galactic center clouds
from the maps of SiO and HNCO towards the Sgr\,A region \citep{Minh05}.

In this work we did not use the emission from the main isotopologue of CS to avoid possible effects due to line opacity.
Instead, we have used the HNCO/$^{13}$CS as a diagnostic ratio which should be basically unaffected by opacity effects \citep{Hutte98}.
As mentioned, the HNCO emission is also optically thin. 

\begin{figure}
\includegraphics[angle=-90,scale=0.7]{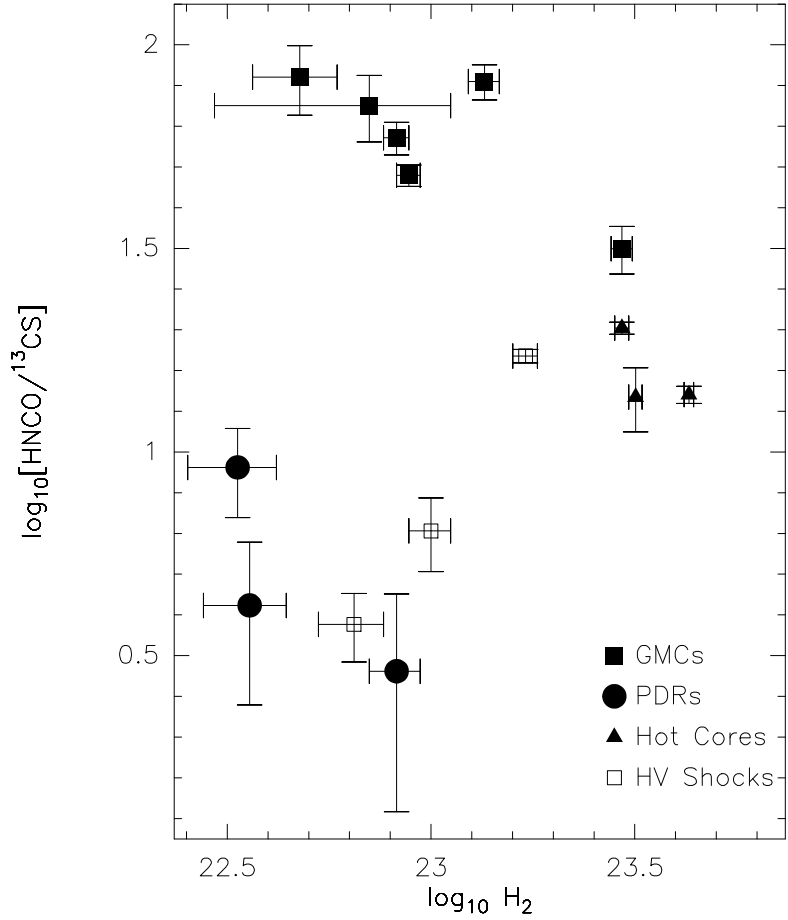}
\caption{Relative abundance ratio HNCO/$^{13}$CS versus H$_2$ column density \citep[$\rm C^{18}O/H_2=1.7\times10^{-7}$,][]{Frerking82}
in logarithmic scale for the sample of sources observed in the Galactic center.
The different symbols represent the PDR clouds, GC giant molecular clouds, hot cores, and sources affected by high velocity shocks.}
\label{fig:vsplot}
\end{figure}

From Fig.~\ref{fig:abundances}, we notice that whenever the relative abundance of HNCO is higher than average, that of $^{13}$CS tends to be lower
than average and viceversa.
In fact, the HNCO/$^{13}$CS or HNCO/C$^{34}$S ratios show much larger variations than the relative abundance variation of either of the molecules involved.
This large differentiation in the HNCO abundance is clearly evidenced in Fig.~\ref{fig:vsplot}, where we show the abundance ratio of HNCO/$^{13}$CS 
versus the H$_2$ column density. The HNCO/$^{13}$CS ratio varies by more than one order of magnitude between different sources.
As for the HNCO/C$^{34}$S ratios we found the same variations in the HNCO/$^{13}$CS ratios.
Most of this variation is due to the HNCO and only to a lesser degree to CS.

From the HNCO/$^{13}$CS abundance ratios in Fig.~\ref{fig:vsplot} we can clearly observe that the sources are clustered in three different categories,
(described in detail below):
a) Galactic center molecular clouds, represented as filled squares, showing moderate H$_2$ column densities with a high HNCO/$^{13}$CS abundance ratio, b) PDRs
shown as filled circles and high velocity shocks as open squares with a low  HNCO/$^{13}$CS abundance ratio, and c) hot cores, filled triangles, with larger H$_2$ column
densities and intermediate HNCO/$^{13}$CS abundance ratios.



\subsection{Typical Galactic center clouds}
\label{sect.GMCs}

The typical Galactic center clouds, hereafter {\it ``giant molecular clouds''} (GMCs), are observed to have sizes of $\sim 30$\,pc, 
densities around $10^{-4}$\,cm$^{-3}$, and kinetic temperatures $\geq80$\,K. 
These sources do not show any traces of recent or ongoing star formation activity \citep{Gusten81,Hutte93,Pintado97}.
We find the largest HNCO/$^{13}$CS abundance ratios towards the set of sources representative of these typical Galactic center 
GMCs (filled squares in Fig.~\ref{fig:vsplot}).
This group of sources includes the offset positions ($20'',100''$) and ($20'',-180''$) in the molecular envelope of Sgr\,B2M, G$+0.24+0.01$, G$-0.11-0.08$, and 
G$+0.83-0.18$.
Though located in very different environments and separated by more than 100$\,$pc within the Galactic center region, all these GMCs share very similar relative abundances of HNCO to CS.
We find a mean HNCO/$^{13}$CS abundance ratio of $68\pm13$ for these sources.

The large abundances of complex molecules and of SiO in these molecular clouds are thought to be due to shocks, which eject these species from grain surfaces.
These shocks must have a sufficiently low velocity to guarantee that the molecules are not dissociated \citep{Hutte98,Pintado01,Requena06}.
Similarly, the large abundance of HNCO observed in this work towards these positions 
can be well explained by gas-phase injection via dust grain erosion by shocks in the molecular clouds across the Galactic center.
Further support to this idea comes from the observations towards six sources common to those observed in this paper which show
enhanced CH$_3$OH/CS abundance ratios by a factor of $\sim$25,
while the 
CS/H$_2$ ratio remains unchanged \citep{Requena06}.

\subsection{Photodissociation regions}
\label{sect.PDRs}

We also observe a group of sources with a relative HNCO/$^{13}$CS abundance ratio smaller by more than an order of magnitude than the values observed
 in the typical GMCs.
This is the case of the observed position in the CND around Sgr\,A$^*$ and the source G$+0.18-0.04$ (Filled dots in Fig.~\ref{fig:vsplot}).
These two sources (hereafter {\it ``PDR clouds''}) are both located in the vicinity of massive star clusters and therefore 
are known to be the best candidates for a chemistry driven by the UV radiation from the stellar clusters.

The Sgr\,A$^*$ observed position is located in the CND, $\sim 1.5$\,pc  (assuming a distance of $\sim8$\,Kpc to the GC) away from the 
nominal position of Sgr\,A$^*$, and it is thus strongly pervaded by the UV radiation from the Central Cluster \citep{Krabbe95}.
G$+0.18-0.04$, in the region known as ``the Sickle'', is strongly irradiated by the OB stars in the Quintuplet Cluster \citep{Figer99,Rodriguez01}.
The difference of a factor of $\sim 3$ in the HNCO/$^{13}$CS abundance ratios observed between the two velocity components at 20 and 70\,km/s$^{-1}$ (the two leftmost 
filled dots in Fig.~\ref{fig:vsplot}) are likely attributed to different degrees of UV irradiation due to the different spatial distribution of each component
\citep{Serabyn91}. 

In extreme conditions with strong UV radiation field in the surroundings of the GC massive star clusters, the photo-resistant CS is expected to 
survive \citep{Goico06}, 
while the easily photodissociated HNCO molecule will quickly disappear from the gas phase as we get close to the source of UV radiation.
In fact, HNCO is observed to be less resistant to the UV radiation than other easily dissociated molecules such as CH$_3$OH \citep{Requena06}, 
also efficiently formed and ejected from dust grains \citep{Millar91,Charnley95}.

In addition to the previous sources, we observe another cloud with also very low HNCO/$^{13}$CS abundance ratio, namely the position G$+0.02-0.02$.
This object is a very compact cloud thought to be the result of the acceleration, heating and compression of the
gas by a series of supernovae shocks \citep{Oka99}.
Irradiation by stellar UV photons is unlikely both due to the absence of free-free radio continuum emission associated with this cloud and the fact that it is too far to be
illuminated by the Central Cluster.
The radiative precursor associated with the J-type shocks resulting from the supernovae events which produces a significant UV radiation field \citep{Shull79} might be a likely 
explanation to the observed low HNCO/$^{13}$CS abundances ratios. In this scenario the dissociation of HNCO by the UV radiation generated in the shock would be
equivalent to that observed in PDRs.

The average HNCO/$^{13}$CS ratio for this group of sources is $5\pm2$ in the PDR clouds, more than one order of magnitude smaller than the average value measured in GC clouds.
However, for the best candidate of a PDR cloud, located in the CND, the measured abundance ratio HNCO/$^{13}$CS is more than
20 times smaller than in the GC clouds.
Our observations therefore indicate that the abundance ratio between the photo-resistant CS molecule and the shock injected fragile HNCO molecule is a 
powerful tool to 
discriminate between molecular material heavily affected by a PDR chemistry (PDR clouds) and that largely shielded from the UV radiation. 

\subsection{Hot cores and supernova interaction}

Finally, we find a third group of sources which shows HNCO/$^{13}$CS abundance ratios below those observed in GMCs but still significantly higher than those in PDR clouds.
We observe in Fig.~\ref{fig:vsplot} that all these sources have H$_2$ column densities (as derived from C$^{18}$O) larger by a factor of 5,
than those in the two groups previously discussed.
These sources are hot molecular cores, namely Sgr\,B2 N, M and S, as well as the sources G$-0.02-0.07$ and Sgr\,B2 M ($-40'',0''$).
The Sgr\,B2 M ($-40'',0''$) position is known to be associated with the ridge of a very recent massive star formation site detected by \citet{deVicente00}.
At this position,
the detection of HC$_3$N vibrationally excited emission clearly shows the presence of hot cores like Sgr\,B2R4 associated with massive protostar like those found in Sgr\,B2M, N and S. The
HNCO/$^{13}$CS ratio in this position should therefore be dominated by the hot cores.

We derive fractional abundances of HNCO of $(5\pm2)\times10^{-9}$ for the hot cores in our sample.
The HNCO abundance observed by \citet{Zinchen00} in massive star forming regions, ranged from $2.5\times10^{-9}$, similar to our measurements, down
to $2\times10^{-10}$.
Our data confirm that HNCO appears to be underabundant towards hot molecular cores and that the largest abundances are not found toward the site of massive star formation.
From the maps of HNCO towards G301.12-0.20 and G270.26+0.83 \citep{Zinchen00}, one observes that the HNCO emission does not peak at 
the IRAS position, though position uncertainties are large.
More evident is the case of the Sgr\,B2 complex \citep{Minh98}, where the emission of HNCO appears to surround the C$^{18}$O emission which peaks toward  
the hot cores.
While the HNCO abundance in Sgr\,B2(M) is $1\times10^{-9}$ \citep[a factor of 7 smaller than our measurements;][]{Minh98}, it reaches values up to $1\times10^{-8}$
in a position 2$'$ north Sgr\,B2(M) ($28''$ away from our $20'',100''$ position).
Therefore the HNCO molecule seems to avoid the hot cores in massive star forming regions and most of its emission is observed to stem from the 
surrounding molecular gas envelope around the hot cores where, similar to that observed in GMCs (Sect~\ref{sect.GMCs}), low velocity shocks would be responsible
for the enhancement of HNCO emission \citep{Flower95}.
There are two exceptions in the sample from \citet{Zinchen00},
namely Orion-KL and Sgr\,A, which show larger abundances of HNCO, $8.7\times10^{-9}$ and $6.3\times10^{-9}$, respectively.
The abundance towards Sgr\,A is in agreement with our findings in the GC clouds, as this refers to a position $65''$ away from our source G$-0.11-0.08$.
For the Orion-KL hot core, the difference in abundance might be related to the evolutionary state of this particular molecular core.

We have also observed that G$-0.02-0.07$ shows a relative low HNCO/$^{13}$CS abundance ratio.
G$-0.02-0.07$, known as the $+50\rm km\,s^{-1}$ cloud \citep{Gusten81}, appears to be surrounding and interacting with Sgr\,A East, a supernova remnant located
just behind the Galactic center \citep[i.e.][]{Zylka90}. Similar to what we observe in G$+0.02-0.02$ (Sect~\ref{sect.PDRs}), the UV field associated with the 
high velocity shocks resulting from the supernovae interaction might be the responsible for the lower HNCO/$^{13}$CS abundance ratio observed in this source.

\subsection{The origin of the large HNCO abundance variation in the GC}

The fractional abundance of HNCO derived from the data presented in this work appear to be fairly constant in the typical Galactic center GMCs.
This uniform abundance is significantly more evident when the HNCO/CS ratio is considered.
As already mentioned above, the most likely origin of gas-phase HNCO is via its ejection from dust grains by low velocity shocks erosion.
Similar uniformity, within a factor of 4, is also found throughout the Galactic center region for the relative abundances
of complex organic species ejected from grains \citep{Requena06,Requena07}.

However, we observe that the HNCO abundances are strongly affected by the extreme physical processes present in hot molecular cores, supernova remnants
and mainly photodissociation regions.
With photodissociation rates almost twice as large as those of CH$_3$OH and slightly larger than those of NH$_3$ \citep{Roberge91}, HNCO seems to be
the most sensitive molecule to the UV radiation fields, even more than NH$_3$.
Thus, while HNCO survives in the gas-phase in the well shielded dense molecular clouds, it is almost completely wiped out in the unshielded regions of 
highly UV-irradiated clouds.
This effect is also observed in the interfaces between supernovae shocks and the molecular environment, where the UV radiation generated in the hot plasma
of the shock front is enough to photoionize and dissociate the molecular gas entering the shock.

This variation is clearly evident in the HNCO/CS ratio, not only due to the higher photodissociation energy of CS, but for the favored production of CS via
ion reactions as a consequence of the mentioned enhanced abundance of S$^+$ in low UV-shielded regions \citep[$A_v < 7$][]{Stern95,Mauers03}.

Additionally, the HNCO abundance is also affected to a lesser extent in dense star forming molecular cores.
As noticed by \cite{Zinchen00}, the HNCO emission towards massive star forming cores does not show the high velocity wings observed in SiO, indicating that,
though HNCO is enhanced by low-velocity shocks, its notoriously underabundant in the presence of high velocity shocks.
Though \cite{Zinchen00} explain the low HNCO abundance as a result of the destruction of O$_2$ (an important molecule in the pathways to gas-phase HNCO formation)
by the high velocity shocks, the low efficiency of HNCO formation in gas-phase as well as its overall homogeneous abundance over the Galactic center region,
seems to point to the destruction of the HNCO molecule itself by fast shocks.
The shock-induced UV field  \citep{Viti02} might account for the dissociation of HNCO in high velocity shocks.

\subsection{From Galactic center to Galaxies}
In the last years it has been recognized that the chemical complexity of the ISM in the nuclei of galaxies can be used to trace the processes which dominate their
heating \citep{Martin06b}. Rarer molecules like SiO, HCO and HOC$^+$, which can be detected in nearby galaxies with different types of nuclear activity has been
suggested to trace heating by PDRs and XDRs from Active Galactic Nuclei (AGN) like NGC1068 \citep{Usero04}, or by shocks in Starburst Galaxies (SB) like 
NGC\,253 \citep{Martin06b}. Unfortunately, these lines are too weak to be detected beyond the more nearby galaxies.

In order to establish a chemical classification of the ISM in the nuclei of galaxies, that can be used out to moderate (z$\sim$1) distances,
and in the future even at higher redshift with ALMA, we need to identify the most suitable molecular tracers of the different dominating chemistries that also 
show strong emission lines.
Diagnostic diagrams of line intensities of abundant molecules like HCN, HNC, HCO$^+$, and CO have been proposed to discriminate between SB and AGN activity
in Galaxies by \citet{Kohno99,Kohno05,Aalto07,Krips07}.
It is claimed that the HCN/HCO$^+$ line intensity ratio versus the HCN/CO line intensity ratio is a good discriminator between SB and AGN dominated nucleus.

The main drawbacks of the strong lines from very abundant molecules with high dipole moment like
HCO$^+$ and HCN is that they can be severely affected by opacity effects \citep{Pintado08} and their
line intensity ratios do not reflect the actual abundance ratios which are directly related to the
chemistry. Further more the dynamic range of line intensity ratios found in different types of galaxies are of only a factor $\lesssim$4 \citep{Kohno07}.

The results from this work point out that the relative abundance of HNCO seems to be one of the best discriminators between the chemistry driven by low 
velocity shocks found in SB like NGC253 and that driven by photodissociation due to the UV radiation from stellar clusters or AGNs.
This molecule has the additional advantage to have transitions near those of C$^{18}$O, allowing a precise determination of the fractional 
abundances, as well as reasonable intensities (of the order of C$^{18}$O) for extragalactic studies.
As an observability comparison with lines such as HCN, if we assume the luminosity of HNCO lines to be $\sim5$ fainter than that of HCN($1-0$)
we estimate a luminosity of $L'_{HNCO}>6\times10^8\,L_\odot$ for ultraluminous infrarred galaxies
\citep[ULIRGs, using the $L'_{HCN}/L_{\rm FIR}$ relation from][]{Gao04}.
For a ULIRG at z=1, we obtain an integrated intensity of $>20\rm\,mJy\,km\,s^{-1}$ observed 80\,GHz.
Assuming a linewidth of 200\,km\,s$^{-1}$, line intensity would be $>0.1$\,mJy.
ALMA would detect this line at a $3\,\sigma$ detection level in 9 hours of time for a $100\,km\,s^{-1}$ resolution.

Additionally, compared to other potentially good tracers of shock driven chemistry such as CH$_3$OH, which has been easily detected towards a number of
starburst galaxies \citep{Henkel97} and recently also towards M\,82 \citep{Martin06a},
HNCO shows much simpler spectral features, which is fundamental when dealing with extragalactic broad 
line profiles with typical linewidths of a few hundred km\,s$^{-1}$, as well as a
higher abundance contrast than CH$_3$OH.

\begin{figure}
\includegraphics[angle=-90,scale=0.7]{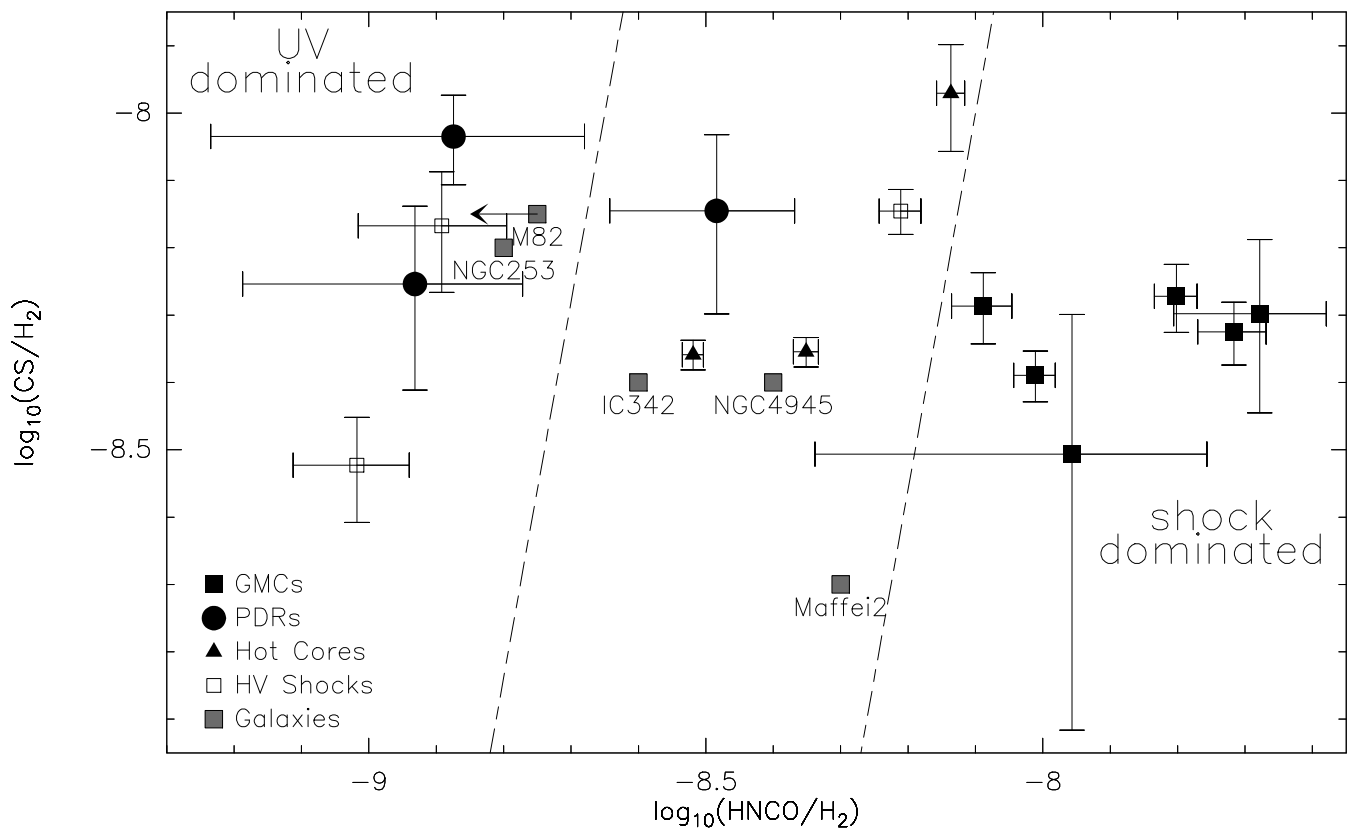}
\caption{Proposed diagnostic diagram. The abundance relative to H$_2$ of CS versus that of HNCO in logarithmic scale for the sample of sources observed in the Galactic center.
Symbols are as in Fig.~\ref{fig:vsplot}. The diagram has the potential to discriminate between the UV (PDR, upper left corner) and shocks (GMCs, lower right)
dominating the ISM in  the nuclei of galaxies.
Available data towards five starburst galaxies are represented by grey filled squares, where the upper HNCO detection limit towards M\,82 is represented with
an arrow \citep[see Table 7 in][for references]{Martin06b}.
Additional details on this plot are explained in the text.}
\label{fig:extravsplot}
\end{figure}

In Fig.~\ref{fig:extravsplot} we present a diagnostic diagram using the relative abundance of CS versus that of HNCO.
We have used the ratio $^{12}$C/$^{13}$C$=20$ \citep{Wilson94} to calculate the CS abundances in the GC sources.
The ratios show a very large dynamic range of up to 1.5 orders of magnitude.
If we compare the GMCs and PDRs abundances, we observe a significant negative correlation between the HNCO and CS abundances as expected from the different
origin of both molecules.
This diagnostic plot shows a clear differentiation between sources with different types of chemistry.
This abundance ratio diagram could be used as diagnostic to discriminate in extragalactic sources between the two main physical processes expected to 
be dominating in the chemistry of the ISM in SB galaxies and in AGN.
The dashed lines in Fig.~\ref{fig:extravsplot} have been traced perpendicular to the CS-HNCO correlation observed between GMCs and PDRs to guide the eye
in the discrimination.
AGNs, with a chemistry expected to be dominated by PDRs, would appear in the upper left corner while the non-UV irradiated material will be observed at 
the lower right corner with very high relative abundances of HNCO.

Unfortunately only a few galaxies have been measured in HNCO and all those measured correspond to starburst galaxies.
We have added to the HNCO diagnostic diagram, the abundance ratios derive for NGC\,253, M\,82, IC\,342, NGC\,4945, and Maffei\,2.
These data clearly show the differences in the heating and the chemistry of the molecular gas in each galaxy, closely related to the state of evolution of their
nuclear starbursts.
The upper limit to the HNCO abundance of M\,82 clearly locate this galaxy in the region of UV dominated material.
This agrees well with the finding from other molecules like HCO towards this galaxy \citep{Burillo02} suggesting M\,82 as the best case of late stage
PDR dominated galaxy \citep{Martin06b}.
Surprisingly, the early state starbursts in NGC\,253, assumed to be mainly dominated by shocks \citep{Martin03,Martin05,Martin06b},
shows abundances of HNCO within the range of the PDR region of the diagram, which suggest that a strong UV field is also affecting an important fraction
of its molecular gas.
Galaxies such as IC\,342 and NGC\,4945 show intermediate abundances suggesting an intermediate state of evolution as found in previous molecular studies
\citep{Wang04,Meier05,Martin06b}, while Maffei\,2 seems to be a source where PDRs have a lesser impact on the chemistry.

This diagnostic diagram should be considered complementary to similar efforts being carried out with HCN, HCO$^+$ and HNC and should be checked 
through observations of the HNCO and CS emission in a sample of selected  galactic nuclei with different types of activity.
These observations are required to test the potential of the proposed CS vs HNCO diagnostic diagram.


\section{Conclusions}
We have studied the HNCO, $^{13}$CS, C$^{34}$S, and C$^{18}$O emission towards 13 positions in the Central Molecular
Zone in the Galactic center region.
Column densities and excitation temperatures have been derived for each position.
These set of sources represent a sample of the main different chemical processes affecting the molecular material.

With variations of up to a factor of 20 in abundance, the molecule of HNCO is shown to be one of the most contrasted species among the observed sources.
The derived HNCO/$^{13}$CS abundance ratio shows particularly high contrast of up to a factor of 30 between sources.
This ratio allows the grouping of the observed sources in three categories: 1) Giant Molecular Clouds showing a high HNCO/$^{13}$CS
ratio, 2) Photo dissociated regions and high velocity shocks interactions with the lowest observed ratios, and 3) hot molecular cores with intermediate
HNCO/$^{13}$CS ratios but with significantly higher H$_2$ column densities.

The negative correlation between the CS and HNCO abundances relative to H$_2$ observed in these groups of sources support the idea of HNCO being ejected from
grain mantles by low velocity shocks while CS formation is favored in PDRs environments via ion reactions.

We have used the results from the Galactic center sources as templates for the observed chemistry in extragalactic nuclei.
Thus, we compared the observed abundances in the Galactic center with the available data on five starburst galaxies.
A similarly large variation in the HNCO abundance is observed with sources such as M\,82 and NGC\,253 showing abundances similar to the galactic PDRs while,
on the other end, Maffei\,2 appears to be more similar to the Galactic center GMCs.
The CS vs HNCO abundance diagram is suggested as a powerful tool to study the chemical differentiation in the nuclei of galaxies.

\acknowledgments
This work has been partially supported by the Spanish Ministerio de Educaci\'on y Ciencia under projects ESP2007-65812-C02-01 , ESP2004-00665,
by the Comunidad de Madrid Government under PRICIT project S-0505/ESP-0237 (ASTROCAM),
and by DGI Grant AYA 2005-07516-C02-01.

{\it Facilities:} \facility{IRAM 30m}

\clearpage

\begin{table}
\begin{center}
\caption{Coordinates and location of observed sources}
\label{tab:sources}
\begin{tabular}{llll}
\tableline
\tableline
Source           &  $\alpha_{J2000}$               &   $\delta_{J2000}$              &   Region        \\
\tableline
Sgr~B2M                 &   $17^{\rm h}47^{\rm m}20\fs4$  & $-28\arcdeg23\arcmin07\arcsec$  &  Sgr B          \\
Sgr~B2M($20'',100''$)   &   $17^{\rm h}47^{\rm m}21\fs9$  & $-28\arcdeg21\arcmin27\arcsec$ &  Sgr B          \\
Sgr~B2M($-40'',0''$)    &   $17^{\rm h}47^{\rm m}17\fs4$  & $-28\arcdeg23\arcmin07\arcsec$ &  Sgr B          \\
Sgr~B2M($20'',-180''$)  &   $17^{\rm h}47^{\rm m}21\fs9$  & $-28\arcdeg26\arcmin07\arcsec$  &  Sgr B          \\
Sgr~B2S                 &   $17^{\rm h}47^{\rm m}20\fs5$  & $-28\arcdeg23\arcmin45\arcsec$  &  Sgr B          \\
Sgr~B2N                 &   $17^{\rm h}47^{\rm m}20\fs3$  & $-28\arcdeg22\arcmin19\arcsec$  &  Sgr B          \\
G$+0.24+0.01$           &   $17^{\rm h}46^{\rm m}09\fs8$  & $-28\arcdeg43\arcmin42\arcsec$  &  Dust Ridge     \\
G$-0.11-0.08$           &   $17^{\rm h}45^{\rm m}38\fs8$  & $-29\arcdeg04\arcmin05\arcsec$  &  Sgr A          \\
G$+0.83-0.18$           &   $17^{\rm h}48^{\rm m}16\fs6$  & $-28\arcdeg19\arcmin17\arcsec$  &  Sgr B          \\
G$-0.02-0.07$           &   $17^{\rm h}45^{\rm m}50\fs6$  & $-28\arcdeg59\arcmin09\arcsec$  &  Sgr A          \\
Sgr~A$^*$\,\tablenotemark{a} &  $17^{\rm h}45^{\rm m}37\fs7$  & $-29\arcdeg00\arcmin58\arcsec$  &  Sgr A        \\
G$+0.02-0.02$           &  $17^{\rm h}45^{\rm m}42\fs8$  & $-28\arcdeg55\arcmin51\arcsec$  &  Sgr A          \\ 
G$+0.18-0.04$           &   $17^{\rm h}46^{\rm m}11\fs3$  & $-28\arcdeg48\arcmin22\arcsec$  &  The Sickle     \\
\tableline
\end{tabular}
\tablenotetext{a}{The position observed in this work correspond to the offset $(-30'',-30'')$ relative to the nominal position of Sgr\,A$^*$.}
\end{center}
\end{table}

\begin{deluxetable}{l c cc cc c}
\tabletypesize{\scriptsize}
\tablecaption{Integrated intensity derived for the observed transitions of C$^{18}$O, C$^{34}$S and $^{13}$CS\label{tab:lines}}
\tablewidth{0pt}
\tablehead{
            &     V$_{\rm LSR}$&\multicolumn{2}{c}{C$^{18}$O}    & \multicolumn{2}{c}{C$^{34}$S}   & $^{13}$CS  \\
            &     (km s$^{-1}$)	&$1-0$          &  $2-1$          &  $3-2$           &  $5-4$       &   $3-2$    \\
SOURCE      &   		&109.782          &  219.560        &   144.617        &   241.016    &   138.739  }
\startdata
{\bf Sgr B2M}   & 60		& 70 (1)   & 128 (2)         &  56 (2)          & 50 (10)      & 40 (1)      \\
($20'',100''$)  & 67		& 27 (1)   & 32 (1)          & 14.3 (0.8)       & 6.6(0.9)     & 8.6(0.5)    \\
($-40'',0''$)   &54		& 7 (1)    & 14 (2)          & 25 (1)           & 7.2 (0.6)    & 8.4 (1.2)   \\
                &71$^*$		& 48 (1)   & 72 (3)	     &  ...		&  ...         & 12.1( 1.7)  \\
($20'',-180''$) &50		& 9.7 (0.6)& 12 (3)	     & 5.2 (0.6)	&  $<0.6$      & 3.4 (0.5)   \\
                &65$^*$		& 6.7 (0.8)& 11 (4)	     &  ...		&  ...         &  ...	     \\
{\bf Sgr B2S}   &61		& 52 (1)   & 87 (1)	     & 27.9 (0.7)	& 13.5 (1.5)   & 17.6 (0.6)  \\
                &92$^*$		& 4.8 (0.8)& 4.4 (0.7)       &  ...		&  ...         &  ...	     \\
{\bf Sgr B2N}   &63		& 65 (2)   & 81 (1)	     & 21 (1)		&  ...         & 48 (8)      \\
{\bf Sgr A$^*$} &20		& 15.8 (0.6)& 24 (3)	      & 16.0 (1.3)	 & 8 (2)	& 10.4 (0.6)  \\
{$\rm \bf G+0.02-0.02$}         &$-13$		& 19.7 (0.4)& 26 (3)	     & 8.6 (0.6)	& 1.5 (0.8)    & 4.0 (0.4)   \\
                &90		& 11.5 (0.3)& 21 (3) 	& 7.4 (0.6)	   & 3.8 (0.8)    & 6.0 (0.5)	\\
{$\rm \bf G+0.24+0.01$}    &34	& 17.6 (1.0)& 22.8 (1.7)     & 7.9 (0.6)        & $<1.6$       & 4.8 (0.2)   \\
{$\rm \bf G-0.11-0.08$}    &17	& 14.9 (0.5)& 23.0 (1.0)     & 7.8 (0.3)        & 2.0 (0.5)    & 6.1 (0.4)   \\
                           &54$^*$	& 4.2 (0.7) & 4.4 (1.1)      &  ...             &  ...         &  ...        \\
{$\rm \bf G+0.83-0.18$}    &38	& 16.0 (0.5)& 15.4 (0.6)     & 2.8 (0.5)        & 1.5 (0.2)    & 3.0 (0.3)   \\
{$\rm \bf G-0.02-0.07$}    &51	& 32.5 (1.2)& 46.8 (1.5)     & 25.5 (0.6)       & 11.7 (0.7)   & 16.6 (0.5)  \\
{$\rm \bf G+0.18-0.04$}    &21	& 6.9 (0.6) & 9.7 (1.8)      & $<1.1$           & $<0.8$       & 2.6 (0.6)   \\
                           &72		& 6.3 (0.6) & 8.5 (1.8)      & $<2.2$           & $<1.6$       & 3.4 (0.8)   \\
\enddata
\tablecomments{Integrated intensities in $T_A^*$ scale (K\,km\,s$^{-1}$). Quantum numbers and frequencies (GHz) are given for each transition.
Different entries for a source indicate resolved velocity components. The velocity components with $^*$ do not show HNCO emission.}
\end{deluxetable}

\begin{deluxetable}{l c c c c c c c c c}
\tabletypesize{\scriptsize}
\tablecaption{Integrated intensity derived for the observed transitions of HNCO\label{tab:lines2}}
\tablewidth{0pt}
\tablehead{
                         &  $5_{0,5}-4_{0,4}$    &  $6_{1,6}-5_{1,5}$ & $6_{0,6}-5_{0,5}$ & $6_{1,5}-5_{1,4}$ & $7_{1,7}-6_{1,6}$ & $7_{0,7}-6_{0,6}$ & $7_{1,6}-6_{1,5}$  \\
SOURCE                   &   109.905             &   131.394          &   131.885         &   132.356         &   153.291         &   153.865         &   154.414          }
\startdata
{\bf Sgr~B2M}            &    64.3 (1.4)         &    13.7 (0.8)      &   76.8 (0.8)      &  11.3 (0.9)       & 18.0 (0.8)        &  72.3 (1.4)       & 18.9 (1.6)         \\
$(20'',100'')$           &   227.2 (0.8)         &     6.2 (1.7)      &   226.0 (1.2)     &  4.5 (1.0)        &  8.1 (0.4)        &  160 (3)          &  4.2 (0.4)         \\
$(-40'',0'')$            &   204.4 (0.5)         &    14.0 (1.1)      &   201.1 (1.4)     &  10.4 (1.4)       &  14.2 (0.8)       &  194.4  (1.3)     &  11.3 (0.8)        \\
$(20'',-180'')$          &    86.2 (0.2)         &     1.0 (0.3)      &    75.2 (1.7)     &  1.7 (0.3)        &  3.3 (0.3)        &  70.4 (0.5)       &  ...               \\
{\bf Sgr~B2N}            &     132 (5)           &      20 (2)        &    112 (3)        & 18.4 (1.6)        &  65.6 (1.6)       &   111 (4)         &    36 (5)          \\
{\bf Sgr~B2S}            &    95.2 (0.4)         &       6 (2)        &   120.6 (1.0)     &  6.4 (0.8)        &  9.9 (0.4)        &  107 (2)          &  10.8 (0.5)        \\
{\bf Sgr~A$^*$}          &     5.4 (0.6)         &       ...          &   6.9 (1.4)       &    ...            &   ...             &  7.2 (0.7)        &  ...               \\
{$\bf G+0.02-0.02$}      &     8.0 (0.3)         &       ...          &   9.3 (0.6)       &    ...            &   ...             &  3.3 (0.3)        &  ...               \\
                         &     6.8 (0.3)         &                    &   8.1 (0.6)       &    ...            &   ...             &  3.2 \tablenotemark{a} &  ...               \\
{$\bf G+0.24+0.01$}      &    76.8 (0.2)         &       ...          &   78.3 (0.8)      &    ...            &  2.1 (0.3)        &  45 (2)           &  ...               \\
{$\bf G-0.11-0.08$}      &   116.9 (0.2)         &       ...          &  112.0 (0.9)      &    ...            &  1.8 (0.2)        &  76 (2)           &  0.6 (0.2)         \\
{$\bf G+0.83-0.18$}      &    64.3 (0.3)         &       ...          &   45.8 (0.6)      &    ...            &   ...             &  26.1 (0.8)       &  0.6 (0.2)         \\
{$\bf G-0.02-0.07$}      &    93.9 (0.2)         &       ...          &   89.8 (1.4)      &  1.5 (0.2)        &   ...             &  56 (2)           &  ...               \\
{$\bf G+0.18-0.04$}      &     1.6 (0.2)         &       ...          &    ...            &    ...            &   ...             &  2.3 (0.3)        &  ...               \\
                         &     7.4 (0.2)         &       ...          &   3.1 (0.8)       &    ...            &   ...             &  2.0 (0.2)        &  ...               \\
\tableline
                         & $8_{1,8}-7_{1,7}$     & $10_{2,k'}-9_{2,k'}$   & $10_{0,10}-9_{0,9}$    & $11_{1,11}-10_{1,10}$   & $11_{1,10}-10_{1,9}$ & $12_{2,k'}-11_{2,k'}$ & $12_{0,12}-11_{0,11}$   \\
                         &   175.189             &   219.733              &   219.798              &   240.876               &   242.639            &   263.672             &   263.748               \\
\tableline
{\bf Sgr~B2M}            &   31 (5)              &   7.1 (1.2)            &  40 (2)                &    16 (2)               &  43.5 (1.8)          &   22 (9)              & 57.3 (1.5)              \\
$(20'',100'')$           &  6.8 (0.4)            &  ...                   &  48 (4)                &    ...                  &  0.9 (0.3)           &   ...                 & 16.4 (0.6)              \\
$(-40'',0'')$            &  14.1 (0.7)           &  4.1 (0.5)             &  94.1 (1.3)            &    6.6 (0.6)            &  4.7 (0.2)           &   ...                 & 44.5 (1.1)              \\
$(20'',-180'')$          &  1.5 (0.1)            &  ...                   &  17.3 (0.5)            &    ...                  &  ...                 &   ...                 & 14.4 (1.1)              \\
{\bf Sgr B2N}            &    49 (2)             &    22 (4)              &    45 (2)              &    66 (6)               &   22 (3)             &    49 (4)             & 81.2 (1.7)              \\
{\bf Sgr B2S}            &  11.0 (0.3)           &  1.9 (1.0)             &  63.2 (1.0)            &    7.5 (0.6)            &  9.7 (1.1)           &   1.6 (0.5)           & 43 (3)                  \\
{$\bf G+0.24+0.01$}      &  1.3 (0.3)            &  ...                   &  10.6 (0.6)            &   ...                   &  ...                 &   ...                 & ...                     \\
{$\bf G-0.11-0.08$}      &  1.2 (0.3)            &  ...                   &  13.8 (0.6)            &   ...                   &  ...                 &   ...                 & ...                     \\
{$\bf G+0.83-0.18$}      &  ...                  &  ...                   &  3.3 (0.5)             &   ...                   &  ...                 &   ...                 & ...                     \\
{$\bf G-0.02-0.07$}      &  2.8 (0.3)            &  ...                   &  14.8 (1.2)            &   ...                   &  ...                 &   ...                 & ...                     \\
\enddata             
\tablecomments{See Fig.~\ref{tab:lines}.}
\tablenotetext{a}{Estimated contamination due to low velocity CH$_3$C$_2$H emission subtracted from fitted profile.}
\end{deluxetable}

 \begin{table}
\setlength{\tabcolsep}{0.03in}
\begin{center}
\caption{Derived column densities and rotational temperatures\label{tab:coldens}}
\begin{tabular}{lcccccccc}
\tableline
\tableline
Source                  &         \multicolumn{2}{c}{C$^{18}$O}            &    \multicolumn{2}{c}{C$^{34}$S}                   &    $^{13}$CS \tablenotemark{a} &        \multicolumn{2}{c}{HNCO}                  \\ 
                        &($\times10^{15}\rm{cm}^{-2}$) & (K)               &($\times10^{13}\rm{cm}^{-2}$) & (K)                 &($\times10^{13}\rm{cm}^{-2}$)   &($\times10^{14}\rm{cm}^{-2}$) & (K)               \\ 
\tableline                                                                                                                                             %
\footnotesize Sgr B2M                 &     73(2)                    &     13.4(0.3)     &  19(4)                       &  18(3)              &    9.4(0.4)                    &  13.0(0.3)                   &    39(0.4)   \\
\footnotesize Sgr B2M($20'',100''$)   &     23(2)                    &      8.7(0.4)     &  5.0(0.7)                    &  12.0(1.0)          &    3.2(0.2)                    &  26(2)                       &   14.8(0.1)    \\ 
\footnotesize Sgr B2M($-40'',0''$)    &     50(3)                    &     10.7(0.5)     &  9.1(0.9)                    &  9.2(0.4)           &    7.6(0.8)                    &  24(2)                       &    19.0(0.1)   \\ 
\footnotesize Sgr B2M($20'',-180'')$  &     8.1(1.9)                 &       11(3)       &  1.9(0.0)                    & 10\tablenotemark{a} &    1.2(0.2)                    &  10(1)                       &     14.8(0.1)  \\ 
\footnotesize Sgr B2N                 &     54(2)                    &      9.0(0.2)     &  7.5(0.4)                    & 10\tablenotemark{a} &    17(3)                       &  23.2(0.7)                   &     55.6(0.9)  \\
\footnotesize Sgr B2S                 &     50(2)                    &     12.2(0.3)     &  9.4(1.0)                    &  11.9(0.8)          &    6.5(0.2)                    &  13.1(0.2)                   &    22.6(0.1)   \\ 
\footnotesize G$+0.24+0.01$           &     15(1)                    &      9.4(0.8)     &  3.5(0.2)                    & 10\tablenotemark{a} &    1.8(0.1)                    &  8.6(0.2)                    &    13.9(0.2)   \\ 
\footnotesize G$-0.11-0.08$           &     14(1)                    &     11.0(0.7)     &  3.0(0.6)                    &  8.6(0.9)           &    2.2(0.2)                    &  13.0(0.2)                   &    12.6(0.1)   \\ 
\footnotesize G$+0.83-0.18$           &     12(7)                    &      7.4(0.3)     &  1.1(0.3)                    &  12.4(1.5)          &    1.1(0.2)                    &  7.8(0.3)                    &    9.1(0.1)    \\ 
\footnotesize G$-0.02-0.07$           &     29(2)                    &     10.3(0.5)     &  8.7(0.5)                    &  11.6(0.4)          &    6.1(0.2)                    &  10.5(0.2)                   &   13.6(0.2)    \\ 
\footnotesize Sgr A$^*$               &     14(2)                    &     10.8(1.3)     &  5.4(1.5)                    &  12(2)              &    3.8(0.2)                    &  1.1(0.6)                    &    35(12)   \\ 
\footnotesize G$+0.02-0.02$           &     17(2)                    &      9.3(0.8)     &  3.4(1.5)                    &  7.6(1.5)           &    1.5(0.2)                    &  0.96(0.15)                  &    10.5(0.7)   \\ 
\footnotesize                         &     11(2)                    &       13(2)       &  2.5(0.6)                    &  12.2(1.7)          &    2.2(0.2)                    &  0.83(0.14)                  &   10.5(0.7)    \\ 
\footnotesize G$+0.18-0.04$           &     6.1(1.4)                 &       10(2)       &  $<0.4$                      & 10\tablenotemark{a} &    1.0(0.2)                    &  0.42(0.16)                  &    46(29)     \\ 
\footnotesize                         &     5.7(1.4)                 &       10(2)       &  $<0.8$                      & 10\tablenotemark{a} &    1.2(0.2)                    &  1.1(0.2)                    &    6.9(0.4)   \\ 
\tableline
\end{tabular}
\tablenotetext{a}{$T_{\rm ex}$ value fixed to 10\,K.}
\end{center}
\end{table}

\clearpage

\clearpage
\thispagestyle{empty}
\setlength{\voffset}{-12mm}

\clearpage
\setlength{\voffset}{0mm}

\clearpage

\end{document}